\documentclass[sigconf]{acmart}
\makeatletter
\renewcommand{\@secfont}{\bfseries\Large}
\makeatother

\usepackage{enumitem}

\AtBeginDocument{%
  }

\copyrightyear{2025}

\begin{document}
\title[Facilitating Accessible Visual Media Exploration through AI-Powered Interactive Storytelling]{Facilitating Visual Media Exploration for Blind and Low Vision Users through AI-Powered Interactive Storytelling}

\author{Shuchang Xu}
\email{sxuby@connect.ust.hk}
\orcid{0000-0002-7642-9044}
\affiliation{
  \department{Computer Science and Engineering}
  \institution{The Hong Kong University of Science and Technology}
  \city{Hong Kong}
  \country{China}
}

\begin{abstract}
Empowering blind and low vision (BLV) users to explore visual media improves content comprehension, strengthens user agency, and fulfills diverse information needs. 
However, most existing tools separate exploration from the main narration, which disrupts the narrative flow, increases cognitive load, and limits deep engagement with visual media. 
To address these challenges, my PhD research introduces the paradigm of \textit{AI-powered interactive storytelling}, 
which leverages AI to generate interactive narratives, enabling BLV users to explore visual media within a coherent storytelling experience. 
I have operationalized this paradigm through three techniques: 
(1) \textit{Hierarchical Narrative}, which supports photo-collection exploration at different levels of detail; 
(2) \textit{Parallel Narrative}, which provides seamless access to time-synced video comments; and 
(3) \textit{Branching Narrative}, which enables immersive navigation of 360° videos. 
Together, these techniques demonstrate that AI-powered interactive storytelling can effectively balance user agency with narrative coherence across diverse media formats. 
My future work will advance this paradigm by enabling more personalized and expressive storytelling experiences for BLV audiences.

\end{abstract}



\begin{teaserfigure}
  \includegraphics[width=\textwidth]{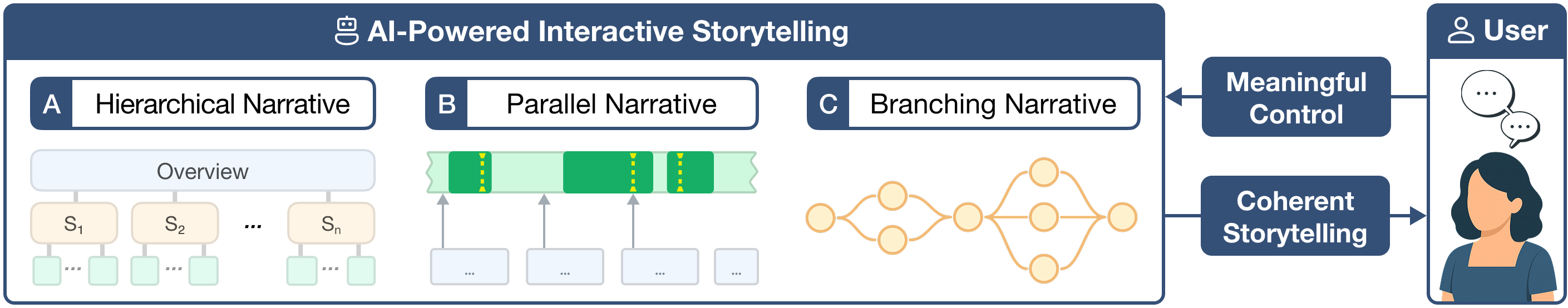}
  \caption{My PhD research explores the paradigm of AI-powered interactive storytelling, which leverages AI to generate interactive narratives, enabling blind and low vision (BLV) users to explore visual media in a coherent storytelling experience. I have operationalized this paradigm through three techniques: (A) \textit{Hierarchical Narrative}, which supports photo-collection exploration at different levels of detail \cite{xu2024memory}; (B) \textit{Parallel Narrative}, which provides seamless access to time-synced user comments in videos \cite{xu2025danmu}; and (C) \textit{Branching Narrative}, which enables immersive navigation of 360° videos \cite{xu2025branch}. These techniques demonstrate the effectiveness of AI-powered interactive storytelling in delivering engaging media experiences for BLV users.}
  \Description{}
  \label{fig:teaser}
\end{teaserfigure}

\maketitle

\section{Introduction}

\begin{figure*}[!t]
    \centering
    \includegraphics[width=\linewidth]{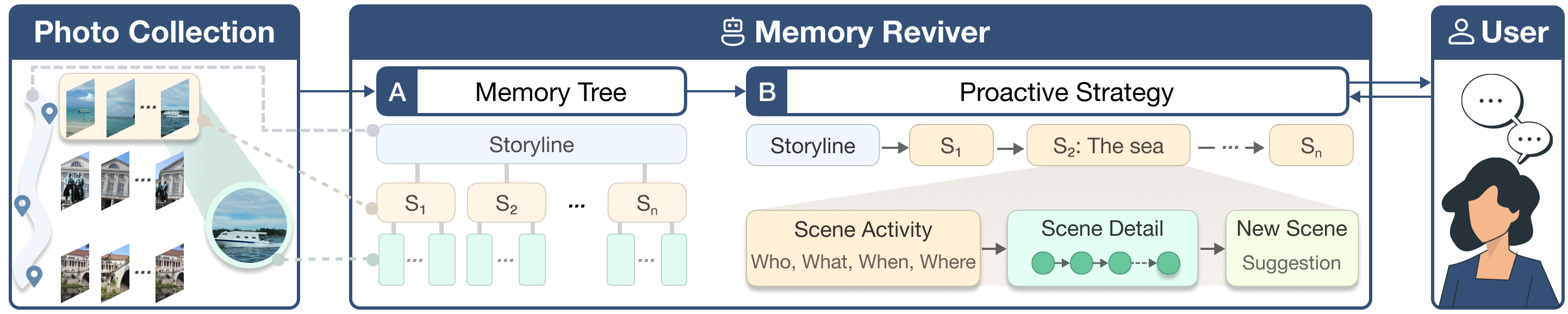}
    \caption{Memory Reviver is a proactive chatbot that guides BLV users to reminisce with a photo collection. It employs a \textit{Hierarchical Narrative}---proactively delivering information at varying levels of detail---to support in-depth reminiscence. Memory Reviver incorporates two core features: (A) a \emph{Memory Tree}, which uses a hierarchical structure to organize information in a photo collection; and (B) a \emph{Proactive Strategy}, which actively delivers information to users at proper conversation rounds.}
    \label{fig:memory_reviver}
\end{figure*}

Visual media (e.g., images and videos) plays a central role in modern communication, education, and entertainment \cite{liu2021makes,jiang2024s}. 
For blind and low vision (BLV) users, access to visual media primarily relies on textual descriptions, such as image captions \cite{stangl2021going,huh2023genassist} and audio descriptions \cite{amy2020rescribe,li2025videoa11y}. While essential, these static and one-size-fits-all descriptions often fail to support deep comprehension or address the diverse needs of BLV audiences \cite{stangl2021going,jiang2024s}. 
In response, recent research has developed systems that enable interactive exploration of visual media \cite{van2024making,peng2021slidecho,ning2024spica,lee2022imageexplorer,stangl2023potential}. These systems allow BLV users to access supplementary descriptions via screen readers \cite{van2024making,huh2023genassist,peng2021slidecho}, explore spatial layouts through touch interfaces \cite{ning2024spica,lee2022imageexplorer,nair_imageassist_2023}, and retrieve visual details using natural language queries \cite{stangl2023potential,antol2015vqa, bigham2010vizwiz}.

However, most existing systems treat exploration as a \textit{separate process} from the main narration \cite{cheema2025describe, ning2024spica, chang2022omniscribe}, which presents two significant challenges. First, users are required to pause the main narration for exploration, which disrupts the narrative flow and diminishes user immersion \cite{chang2022omniscribe,cheema2025describe}. Second, since most systems rely on users to initiate exploration, BLV users must continually decide when and how to seek additional information. This decision-making process increases cognitive load \cite{chang2022omniscribe,cheema2025describe}, heightens a fear of missing out \cite{cheema2025describe}, and hinders deep engagement with visual media.

To address these challenges, my PhD research introduces the paradigm of \textit{AI-powered interactive storytelling} (see Figure~\ref{fig:teaser}), which leverages AI to generate interactive narratives, enabling BLV users to explore visual media within a coherent storytelling experience. 
Achieving this vision relies on two core components: (1) the \emph{narrative structure}, which organizes visual content into an interactive storyline; and 
(2) the \emph{interaction mechanism}, which enables meaningful user control while preserving narrative coherence. 
By effectively coordinating both components, this paradigm balances user agency with narrative continuity, ultimately delivering interactive and engaging media experiences for BLV audiences.

My past work has operationalized this paradigm through three techniques, each tailored to a distinct form of visual media. 
First, I proposed \textbf{Hierarchical Narrative} for photo collections \cite{xu2024memory}, 
which organizes information into a hierarchical structure and proactively presents content at varying levels of detail to deliver a coherent storytelling experience. 
Second, I developed \textbf{Parallel Narrative} for time-synced video comments \cite{xu2025danmu}, 
which seamlessly integrates user comments into the video as part of a coherent narrative, enabling users to enjoy both content simultaneously. 
Third, I designed \textbf{Branching Narrative} for 360° videos \cite{xu2025branch}, which constructs diverse viewing paths, allowing users to choose their own perspectives within an immersive experience. 
Together, these works demonstrate the effectiveness of AI-powered interactive storytelling across visual media that vary in both \textit{temporality} (static images \cite{xu2024memory}  v.s. dynamic videos \cite{xu2025danmu,xu2025branch}) and \textit{spatiality} (2D content \cite{xu2024memory,xu2025danmu} v.s. panoramas \cite{xu2025branch}). 
In future work, I aim to advance this paradigm by enabling more personalized narrative experiences (e.g., narration customization) and exploring more expressive storytelling modalities (e.g., spatial audio) for BLV audiences.

\section{Research Methods and Contributions}

My research develops AI-powered interactive storytelling systems through a three-phase methodology: 

(1) \textit{Design}: I conduct interviews and co-viewing sessions with BLV users to derive core design requirements; 

(2) \textit{Development}: I translate these requirements into AI-driven computational pipelines and interaction techniques that enable interactive storytelling for BLV users; and 

(3) \textit{Evaluation}: I conduct mixed-method user studies, using both quantitative metrics and qualitative results for system evaluation.

Through this approach, I have developed three systems tailored to different forms of visual media: 
photo collections (Sec.~\ref{sec:memory_reviver}), 
time-synced video comments (Sec.~\ref{sec:danmuA11y}), and 
360° videos (Sec.~\ref{sec:branch_explorer}).

\begin{figure*}[!th]
    \centering
    \includegraphics[width=\linewidth]{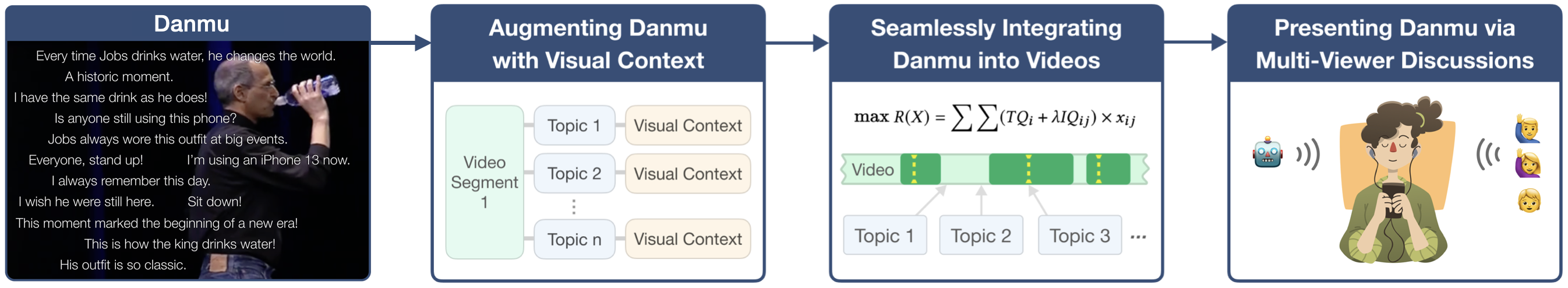}
    \caption{DanmuA11y makes time-synced on-screen video comments (Danmu) accessible to BLV viewers using \textit{Parallel Narrative}. It integrates Danmu comments into the video as part of a coherent storyline, allowing users to engage with both the video and the commentary simultaneously. DanmuA11y incorporates three core features: (1) augmenting Danmu with visual context, (2) seamlessly integrating Danmu into videos, and (3) presenting Danmu via multi-viewer discussions.}
    \label{fig:danmuA11y}
\end{figure*}

\subsection{\textit{Hierarchical Narrative} for Photo-Collection Reminiscence (Memory Reviver)}\label{sec:memory_reviver}

Reminiscing with photo collections offers significant psychological benefits but poses challenges for BLV users. 
Existing tools primarily focus on single-image exploration \cite{lee2022imageexplorer,nair_imageassist_2023}, providing limited support for a comprehensive reminiscence experience. 
To address this limitation, I propose designing a chatbot that enables BLV users to reminisce with photo collections through natural language communication. 
In a formative study, eight BLV participants reminisced with their photo collections by conversing with a GPT-4V-based chatbot, 
revealing two primary challenges: \textit{the scattering of information} and \textit{a lack of proactive guidance}. Firstly, the ``one question, one answer'' communication style led to the information being scattered across multiple rounds, making it hard for users to recall and organize details, preventing them from forming a clear story about their past. 
Secondly, since BLV users could not visually explore new scenes, the chatbot's lack of proactive guidance further challenged users to deeply engage in the reminiscence experience.

To address these challenges, I developed Memory Reviver, a proactive chatbot that guides BLV users to reminisce with a photo collection. 
It employs a \textbf{Hierarchical Narrative} design, which proactively delivers information at varying levels of detail to support in-depth reminiscence. Memory Reviver incorporates two core features (see Figure~\ref{fig:memory_reviver}): 
(1) a \textbf{Memory Tree}, which uses a hierarchical structure to organize the information in a photo collection; and (2) a \textbf{Proactive Strategy}, which actively delivers information to users at proper conversation rounds. Powered by the two features, Memory Reviver delivers a natural conversation flow. It begins the conversation with a clear storyline, helps users recall past activities, enriches their memories by gradually presenting details, and proactively suggests new scenes at detected proper conversation rounds. 
This guided exploration is seamlessly integrated with free-form dialogue, enabling both user agency and narrative coherence throughout the reminiscence experience. 
A comparison study with twelve BLV users demonstrated that Memory Reviver effectively enhanced photo-collection comprehension ($p<.01$), facilitated engaging reminiscence ($p<.01$), and delivered natural conversational experiences ($p<.01$). Overall, Memory Reviver demonstrates the effectiveness of \textit{Hierarchical Narrative} in supporting exploration across different levels of detail while maintaining a coherent narrative experience.

\subsection{\textit{Parallel Narrative} for Time-Synced Video Comments (DanmuA11y)}\label{sec:danmuA11y}

Danmu is a video commenting feature that overlays time-synced user comments onto the video screen (see Figure~\ref{fig:danmuA11y}). It creates a co-watching experience for online viewers \cite{ma2017video,chen2017watching}. However, its visual-centric design poses significant challenges for BLV viewers. Through a formative study with eight BLV participants, I identified three primary challenges that hinder their engagement with Danmu: 
First, the \textit{lack of visual context} makes it difficult for BLV users to understand the discussion topics in the comments. 
Second, the \textit{speech interference} between the Danmu comments and the video prevents users from enjoying both content simultaneously. 
Third, the \textit{disorganization of comments} makes it tedious for BLV viewers to follow audience discussions and socially engage with other viewers.

To address these challenges, I developed DanmuA11y, a system that makes Danmu accessible through \textbf{Parallel Narrative}. It seamlessly integrates Danmu comments into the video to form a coherent narrative, enabling users to enjoy both content simultaneously. 
DanmuA11y achieves this experience through three core features: 
(1) \textbf{Augmenting Danmu with Visual Context}: Danmu comments are supplemented with descriptions of the visual context, allowing BLV viewers to easily grasp the discussion topics; 
(2) \textbf{Seamlessly Integrating Danmu into Videos}: By optimizing the insertion timing of Danmu comments in the video, DanmuA11y enables viewers to enjoy both content without speech overlap; and 
(3) \textbf{Presenting Danmu via Multi-Viewer Discussions}: To create a co-watching experience, DanmuA11y organizes Danmu comments into dialogues and uses spatial audio to simulate the sensation of other viewers conversing around the user. 
DanmuA11y also introduces a simple shake gesture that allows users to retrieve Danmu comments on demand. 
A comparison study with twelve BLV viewers demonstrated that DanmuA11y significantly improved Danmu comprehension ($p<.01$), provided smooth viewing experiences ($p<.01$), and fostered social connections among viewers ($p<.01$). Overall, DanmuA11y demonstrates that \textit{Parallel Narrative} effectively provides seamless access to concurrent narrative threads and fosters an engaging experience for BLV users.

\subsection{\textit{Branching Narrative} for 360° Video Exploration (Branch Explorer)}\label{sec:branch_explorer}

\begin{figure*}[!ht]
    \centering
    \includegraphics[width=\linewidth]{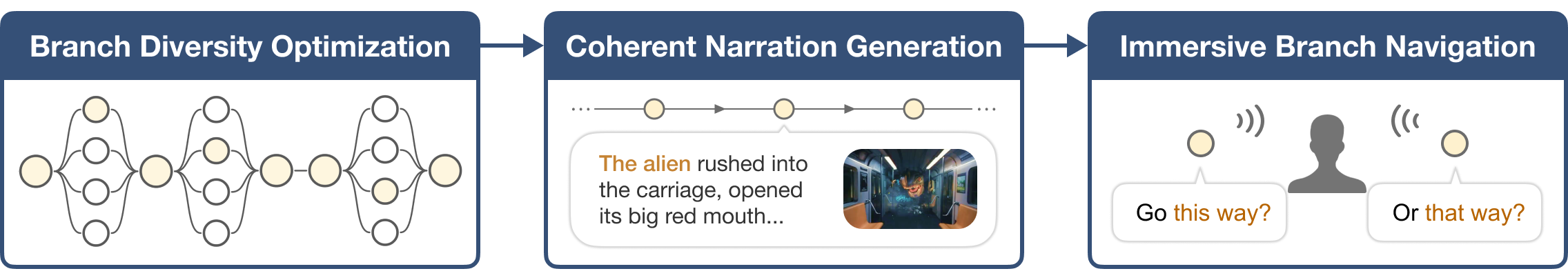}
    \caption{Branch Explorer transforms 360° videos into \textit{Branching Narrative}---stories that dynamically unfold based on viewer choices---to create an engaging experience for BLV users. It incorporates three core modules: (1) \textit{Branch Diversity Optimization}, which employs an optimization approach to curate branches with high spatial, semantic, and social diversity; (2) \textit{Coherent Narration Generation}, which produces coherent narrations and contextual cues to create a smooth narrative flow; and (3) \textit{Immersive Branch Navigation}, which enables flexible navigation between branches, ensuring an immersive viewing experience.}
    \label{fig:branch_explorer}
\end{figure*}

360° video captures a panoramic field of view, enabling users to freely choose their own viewing paths \cite{amy2017uistguidance}. However, BLV users are largely excluded from this interactive experience. To bridge this gap, I propose using \textbf{Branching Narrative} (see Figure~\ref{fig:branch_explorer})—a technique that constructs multiple diverging viewing paths, enabling users to choose their perspectives. 
Through a formative study with eight BLV participants, I identified three key considerations for designing accessible branching narratives: 
(1) \textit{Providing diverse branch options} that vary in spatial, semantic, and social aspects; 
(2) \textit{Ensuring coherent story progression} by optimizing the frequency and timing of choices; and 
(3) \textit{Enabling immersive navigation among branches} by adapting interaction methods to different user contexts.

Guided by these insights, I developed Branch Explorer, a system that transforms 360° videos into branching narratives, enabling interactive and immersive experiences for BLV users. 
Branch Explorer incorporates three modules: 
(1) \textbf{Branch Diversity Optimization}, 
which generates varied narrative paths, allowing users to explore the video from diverse spatial, semantic, and social perspectives; 
(2) \textbf{Coherent Narration Generation}, 
which produces consistent narrations and contextual cues to maintain a smooth narrative flow; and 
(3) \textbf{Immersive Branch Navigation}, 
which leverages subtle guidance and flexible user input to support immersive navigation among narrative paths. 
Powered by these modules, Branch Explorer allows users to engage with diverse storylines, make choices at branching points, and revisit alternative paths during replay. 
A comparison study with twelve BLV participants showed that Branch Explorer significantly enhanced users' sense of agency ($p<.01$), narrative immersion ($p<.01$), and overall engagement ($p<.05$) during 360° video experiences. This work further provides implications for supporting accessible exploration of videos and virtual environments for BLV audiences.

\section{Future Directions}
In future work, I aim to advance AI-powered interactive storytelling by (1) enabling real-time narration generation, and (2) exploring more expressive storytelling possibilities.

\subsection{Real-Time Narration Generation}
My past work relies on pre-extracted narratives to accelerate system responses for real-time interaction. However, such pre-generated content cannot adapt to diverse, real-time user interests---for example, prioritizing a specific character in a film \cite{jiang2024s}. To address this limitation, my future work will explore real-time narration generation that dynamically responds to both the visual content and the viewer's evolving interests. This direction presents both technical and human-factor challenges. On the \textbf{human-factor} side, I will investigate how to support users in expressing their interests more fluidly, using both explicit input (e.g., quick-access menus \cite{natalie2024audio}) and implicit signals (e.g., fast-forwarding or rewinding behavior). 
On the \textbf{technical} side, interactive storytelling should maintain global narrative coherence and integrate seamlessly with the original content (e.g., the video's soundtrack) \cite{amy2020rescribe}. My future work will explore generating coherent, real-time narration by combining local user input, prior narration context, and the media's global structure, ultimately enabling a more adaptive and personalized storytelling experience for BLV audiences.

\subsection{More Expressive Storytelling Possibilities}
My past work has primarily focused on using textual descriptions to enable interactive storytelling. However, text alone has limitations---particularly in conveying spatial information \cite{chheda2025brought} and enhancing immersion \cite{jiang2023beyondAD}. In future work, I plan to explore more expressive storytelling modalities, such as spatial audio \cite{uist2023front_row} and haptic feedback \cite{reinders2025brought}, to create richer and more immersive storytelling experiences for BLV users.

Additionally, while my previous research has focused on images and videos, future work could extend interactive storytelling to more dynamic scenarios such as video games \cite{Surveyor}, virtual reality \cite{collins2023guide}, and real-world environments \cite{uist24worldscribe}. This opens up opportunities to develop personalized storytelling assistants that can adaptively guide BLV users through dynamic, temporally evolving environments.

\bibliographystyle{ACM-Reference-Format}
\bibliography{sample-base}


\begin{thebibliography}{30}


\ifx \showCODEN    \undefined \def \showCODEN     #1{\unskip}     \fi
\ifx \showISBNx    \undefined \def \showISBNx     #1{\unskip}     \fi
\ifx \showISBNxiii \undefined \def \showISBNxiii  #1{\unskip}     \fi
\ifx \showISSN     \undefined \def \showISSN      #1{\unskip}     \fi
\ifx \showLCCN     \undefined \def \showLCCN      #1{\unskip}     \fi
\ifx \shownote     \undefined \def \shownote      #1{#1}          \fi
\ifx \showarticletitle \undefined \def \showarticletitle #1{#1}   \fi
\ifx \showURL      \undefined \def \showURL       {\relax}        \fi
\providecommand\bibfield[2]{#2}
\providecommand\bibinfo[2]{#2}
\providecommand\natexlab[1]{#1}
\providecommand\showeprint[2][]{arXiv:#2}

\bibitem[Antol et~al\mbox{.}(2015)]%
        {antol2015vqa}
\bibfield{author}{\bibinfo{person}{Stanislaw Antol}, \bibinfo{person}{Aishwarya Agrawal}, \bibinfo{person}{Jiasen Lu}, \bibinfo{person}{Margaret Mitchell}, \bibinfo{person}{Dhruv Batra}, \bibinfo{person}{C~Lawrence Zitnick}, {and} \bibinfo{person}{Devi Parikh}.} \bibinfo{year}{2015}\natexlab{}.
\newblock \showarticletitle{Vqa: Visual question answering}. In \bibinfo{booktitle}{\emph{Proceedings of the IEEE international conference on computer vision}}. \bibinfo{pages}{2425--2433}.
\newblock


\bibitem[Bigham et~al\mbox{.}(2010)]%
        {bigham2010vizwiz}
\bibfield{author}{\bibinfo{person}{Jeffrey~P Bigham}, \bibinfo{person}{Chandrika Jayant}, \bibinfo{person}{Hanjie Ji}, \bibinfo{person}{Greg Little}, \bibinfo{person}{Andrew Miller}, \bibinfo{person}{Robert~C Miller}, \bibinfo{person}{Robin Miller}, \bibinfo{person}{Aubrey Tatarowicz}, \bibinfo{person}{Brandyn White}, \bibinfo{person}{Samual White}, {et~al\mbox{.}}} \bibinfo{year}{2010}\natexlab{}.
\newblock \showarticletitle{Vizwiz: nearly real-time answers to visual questions}. In \bibinfo{booktitle}{\emph{Proceedings of the 23nd annual ACM symposium on User interface software and technology}}. \bibinfo{pages}{333--342}.
\newblock


\bibitem[Chang et~al\mbox{.}(2024)]%
        {uist24worldscribe}
\bibfield{author}{\bibinfo{person}{Ruei-Che Chang}, \bibinfo{person}{Yuxuan Liu}, {and} \bibinfo{person}{Anhong Guo}.} \bibinfo{year}{2024}\natexlab{}.
\newblock \showarticletitle{WorldScribe: Towards Context-Aware Live Visual Descriptions}. In \bibinfo{booktitle}{\emph{Proceedings of the 37th Annual ACM Symposium on User Interface Software and Technology}} (Pittsburgh, PA, USA) \emph{(\bibinfo{series}{UIST '24})}. \bibinfo{publisher}{Association for Computing Machinery}, \bibinfo{address}{New York, NY, USA}, Article \bibinfo{articleno}{140}, \bibinfo{numpages}{18}~pages.
\newblock
\showISBNx{9798400706288}
\href{https://doi.org/10.1145/3654777.3676375}{doi:\nolinkurl{10.1145/3654777.3676375}}


\bibitem[Chang et~al\mbox{.}(2022)]%
        {chang2022omniscribe}
\bibfield{author}{\bibinfo{person}{Ruei-Che Chang}, \bibinfo{person}{Chao-Hsien Ting}, \bibinfo{person}{Chia-Sheng Hung}, \bibinfo{person}{Wan-Chen Lee}, \bibinfo{person}{Liang-Jin Chen}, \bibinfo{person}{Yu-Tzu Chao}, \bibinfo{person}{Bing-Yu Chen}, {and} \bibinfo{person}{Anhong Guo}.} \bibinfo{year}{2022}\natexlab{}.
\newblock \showarticletitle{Omniscribe: Authoring immersive audio descriptions for 360 videos}. In \bibinfo{booktitle}{\emph{Proceedings of the 35th Annual ACM Symposium on User Interface Software and Technology}}. \bibinfo{pages}{1--14}.
\newblock


\bibitem[Cheema et~al\mbox{.}(2025)]%
        {cheema2025describe}
\bibfield{author}{\bibinfo{person}{Maryam Cheema}, \bibinfo{person}{Hasti Seifi}, {and} \bibinfo{person}{Pooyan Fazli}.} \bibinfo{year}{2025}\natexlab{}.
\newblock \showarticletitle{Describe Now: User-Driven Audio Description for Blind and Low Vision Individuals}. In \bibinfo{booktitle}{\emph{Proceedings of the 2025 ACM Designing Interactive Systems Conference}}. \bibinfo{pages}{458--474}.
\newblock


\bibitem[Chen et~al\mbox{.}(2017)]%
        {chen2017watching}
\bibfield{author}{\bibinfo{person}{Yue Chen}, \bibinfo{person}{Qin Gao}, {and} \bibinfo{person}{Pei-Luen~Patrick Rau}.} \bibinfo{year}{2017}\natexlab{}.
\newblock \showarticletitle{Watching a movie alone yet together: understanding reasons for watching Danmaku videos}.
\newblock \bibinfo{journal}{\emph{International Journal of Human--Computer Interaction}} \bibinfo{volume}{33}, \bibinfo{number}{9} (\bibinfo{year}{2017}), \bibinfo{pages}{731--743}.
\newblock


\bibitem[Chheda-Kothary et~al\mbox{.}(2025)]%
        {chheda2025brought}
\bibfield{author}{\bibinfo{person}{Arnavi Chheda-Kothary}, \bibinfo{person}{Ather Sharif}, \bibinfo{person}{David~Angel Rios}, {and} \bibinfo{person}{Brian~A Smith}.} \bibinfo{year}{2025}\natexlab{}.
\newblock \showarticletitle{" It Brought Me Joy": Opportunities for Spatial Browsing in Desktop Screen Readers}. In \bibinfo{booktitle}{\emph{Proceedings of the 2025 CHI Conference on Human Factors in Computing Systems}}. \bibinfo{pages}{1--18}.
\newblock


\bibitem[Collins et~al\mbox{.}(2023)]%
        {collins2023guide}
\bibfield{author}{\bibinfo{person}{Jazmin Collins}, \bibinfo{person}{Crescentia Jung}, \bibinfo{person}{Yeonju Jang}, \bibinfo{person}{Danielle Montour}, \bibinfo{person}{Andrea~Stevenson Won}, {and} \bibinfo{person}{Shiri Azenkot}.} \bibinfo{year}{2023}\natexlab{}.
\newblock \showarticletitle{“The Guide Has Your Back”: Exploring How Sighted Guides Can Enhance Accessibility in Social Virtual Reality for Blind and Low Vision People}. In \bibinfo{booktitle}{\emph{Proceedings of the 25th international ACM SIGACCESS conference on computers and accessibility}}. \bibinfo{pages}{1--14}.
\newblock


\bibitem[Huh et~al\mbox{.}(2023)]%
        {huh2023genassist}
\bibfield{author}{\bibinfo{person}{Mina Huh}, \bibinfo{person}{Yi-Hao Peng}, {and} \bibinfo{person}{Amy Pavel}.} \bibinfo{year}{2023}\natexlab{}.
\newblock \showarticletitle{GenAssist: Making image generation accessible}. In \bibinfo{booktitle}{\emph{Proceedings of the 36th Annual ACM Symposium on User Interface Software and Technology}}. \bibinfo{pages}{1--17}.
\newblock


\bibitem[Jain et~al\mbox{.}(2023)]%
        {uist2023front_row}
\bibfield{author}{\bibinfo{person}{Gaurav Jain}, \bibinfo{person}{Basel Hindi}, \bibinfo{person}{Connor Courtien}, \bibinfo{person}{Xin Yi~Therese Xu}, \bibinfo{person}{Conrad Wyrick}, \bibinfo{person}{Michael Malcolm}, {and} \bibinfo{person}{Brian~A Smith}.} \bibinfo{year}{2023}\natexlab{}.
\newblock \showarticletitle{Front Row: Automatically Generating Immersive Audio Representations of Tennis Broadcasts for Blind Viewers}. In \bibinfo{booktitle}{\emph{Proceedings of the 36th Annual ACM Symposium on User Interface Software and Technology}}. \bibinfo{pages}{1--17}.
\newblock


\bibitem[Jiang et~al\mbox{.}(2024)]%
        {jiang2024s}
\bibfield{author}{\bibinfo{person}{Lucy Jiang}, \bibinfo{person}{Crescentia Jung}, \bibinfo{person}{Mahika Phutane}, \bibinfo{person}{Abigale Stangl}, {and} \bibinfo{person}{Shiri Azenkot}.} \bibinfo{year}{2024}\natexlab{}.
\newblock \showarticletitle{“It’s Kind of Context Dependent”: Understanding Blind and Low Vision People’s Video Accessibility Preferences Across Viewing Scenarios}. In \bibinfo{booktitle}{\emph{Proceedings of the 2024 CHI Conference on Human Factors in Computing Systems}}. \bibinfo{pages}{1--20}.
\newblock


\bibitem[Jiang et~al\mbox{.}(2023)]%
        {jiang2023beyondAD}
\bibfield{author}{\bibinfo{person}{Lucy Jiang}, \bibinfo{person}{Mahika Phutane}, {and} \bibinfo{person}{Shiri Azenkot}.} \bibinfo{year}{2023}\natexlab{}.
\newblock \showarticletitle{Beyond Audio Description: Exploring 360° Video Accessibility with Blind and Low Vision Users Through Collaborative Creation}. In \bibinfo{booktitle}{\emph{Proceedings of the 25th International ACM SIGACCESS Conference on Computers and Accessibility}} (New York, NY, USA) \emph{(\bibinfo{series}{ASSETS '23})}. \bibinfo{publisher}{Association for Computing Machinery}, \bibinfo{address}{New York, NY, USA}, Article \bibinfo{articleno}{50}, \bibinfo{numpages}{17}~pages.
\newblock
\showISBNx{9798400702204}
\href{https://doi.org/10.1145/3597638.3608381}{doi:\nolinkurl{10.1145/3597638.3608381}}


\bibitem[Lee et~al\mbox{.}(2022)]%
        {lee2022imageexplorer}
\bibfield{author}{\bibinfo{person}{Jaewook Lee}, \bibinfo{person}{Jaylin Herskovitz}, \bibinfo{person}{Yi-Hao Peng}, {and} \bibinfo{person}{Anhong Guo}.} \bibinfo{year}{2022}\natexlab{}.
\newblock \showarticletitle{ImageExplorer: Multi-layered touch exploration to encourage skepticism towards imperfect AI-generated image captions}. In \bibinfo{booktitle}{\emph{Proceedings of the 2022 CHI Conference on Human Factors in Computing Systems}}. \bibinfo{pages}{1--15}.
\newblock


\bibitem[Li et~al\mbox{.}(2025)]%
        {li2025videoa11y}
\bibfield{author}{\bibinfo{person}{Chaoyu Li}, \bibinfo{person}{Sid Padmanabhuni}, \bibinfo{person}{Maryam~S Cheema}, \bibinfo{person}{Hasti Seifi}, {and} \bibinfo{person}{Pooyan Fazli}.} \bibinfo{year}{2025}\natexlab{}.
\newblock \showarticletitle{Videoa11y: Method and dataset for accessible video description}. In \bibinfo{booktitle}{\emph{Proceedings of the 2025 CHI Conference on Human Factors in Computing Systems}}. \bibinfo{pages}{1--29}.
\newblock


\bibitem[Liu et~al\mbox{.}(2021)]%
        {liu2021makes}
\bibfield{author}{\bibinfo{person}{Xingyu Liu}, \bibinfo{person}{Patrick Carrington}, \bibinfo{person}{Xiang'Anthony' Chen}, {and} \bibinfo{person}{Amy Pavel}.} \bibinfo{year}{2021}\natexlab{}.
\newblock \showarticletitle{What makes videos accessible to blind and visually impaired people?}. In \bibinfo{booktitle}{\emph{Proceedings of the 2021 CHI Conference on Human Factors in Computing Systems}}. \bibinfo{pages}{1--14}.
\newblock


\bibitem[Ma and Cao(2017)]%
        {ma2017video}
\bibfield{author}{\bibinfo{person}{Xiaojuan Ma} {and} \bibinfo{person}{Nan Cao}.} \bibinfo{year}{2017}\natexlab{}.
\newblock \showarticletitle{Video-based evanescent, anonymous, asynchronous social interaction: Motivation and adaption to medium}. In \bibinfo{booktitle}{\emph{Proceedings of the 2017 ACM conference on computer supported cooperative work and social computing}}. \bibinfo{pages}{770--782}.
\newblock


\bibitem[Nair et~al\mbox{.}(2023)]%
        {nair_imageassist_2023}
\bibfield{author}{\bibinfo{person}{Vishnu Nair}, \bibinfo{person}{Hanxiu~'Hazel' Zhu}, {and} \bibinfo{person}{Brian~A. Smith}.} \bibinfo{year}{2023}\natexlab{}.
\newblock \showarticletitle{{ImageAssist}: Tools for Enhancing Touchscreen-Based Image Exploration Systems for Blind and Low Vision Users}. In \bibinfo{booktitle}{\emph{Proceedings of the 2023 {CHI} Conference on Human Factors in Computing Systems}} (Hamburg Germany, 2023-04-19). \bibinfo{publisher}{{ACM}}, \bibinfo{pages}{1--17}.
\newblock
\showISBNx{978-1-4503-9421-5}
\href{https://doi.org/10.1145/3544548.3581302}{doi:\nolinkurl{10.1145/3544548.3581302}}


\bibitem[Nair et~al\mbox{.}(2024)]%
        {Surveyor}
\bibfield{author}{\bibinfo{person}{Vishnu Nair}, \bibinfo{person}{Hanxiu~'Hazel' Zhu}, \bibinfo{person}{Peize Song}, \bibinfo{person}{Jizhong Wang}, {and} \bibinfo{person}{Brian~A. Smith}.} \bibinfo{year}{2024}\natexlab{}.
\newblock \showarticletitle{Surveyor: Facilitating Discovery Within Video Games for Blind and Low Vision Players}. In \bibinfo{booktitle}{\emph{Proceedings of the 2024 CHI Conference on Human Factors in Computing Systems}} (Honolulu, HI, USA) \emph{(\bibinfo{series}{CHI '24})}. \bibinfo{publisher}{Association for Computing Machinery}, \bibinfo{address}{New York, NY, USA}, Article \bibinfo{articleno}{10}, \bibinfo{numpages}{15}~pages.
\newblock
\showISBNx{9798400703300}
\href{https://doi.org/10.1145/3613904.3642615}{doi:\nolinkurl{10.1145/3613904.3642615}}


\bibitem[Natalie et~al\mbox{.}(2024)]%
        {natalie2024audio}
\bibfield{author}{\bibinfo{person}{Rosiana Natalie}, \bibinfo{person}{Ruei-Che Chang}, \bibinfo{person}{Smitha Sheshadri}, \bibinfo{person}{Anhong Guo}, {and} \bibinfo{person}{Kotaro Hara}.} \bibinfo{year}{2024}\natexlab{}.
\newblock \showarticletitle{Audio description customization}. In \bibinfo{booktitle}{\emph{Proceedings of the 26th International ACM SIGACCESS Conference on Computers and Accessibility}}. \bibinfo{pages}{1--19}.
\newblock


\bibitem[Ning et~al\mbox{.}(2024)]%
        {ning2024spica}
\bibfield{author}{\bibinfo{person}{Zheng Ning}, \bibinfo{person}{Brianna~L Wimer}, \bibinfo{person}{Kaiwen Jiang}, \bibinfo{person}{Keyi Chen}, \bibinfo{person}{Jerrick Ban}, \bibinfo{person}{Yapeng Tian}, \bibinfo{person}{Yuhang Zhao}, {and} \bibinfo{person}{Toby Jia-Jun Li}.} \bibinfo{year}{2024}\natexlab{}.
\newblock \showarticletitle{SPICA: Interactive Video Content Exploration through Augmented Audio Descriptions for Blind or Low-Vision Viewers}. In \bibinfo{booktitle}{\emph{Proceedings of the 2024 CHI Conference on Human Factors in Computing Systems}} (Honolulu, HI, USA) \emph{(\bibinfo{series}{CHI '24})}. \bibinfo{publisher}{Association for Computing Machinery}, \bibinfo{address}{New York, NY, USA}, Article \bibinfo{articleno}{902}, \bibinfo{numpages}{18}~pages.
\newblock
\showISBNx{9798400703300}
\href{https://doi.org/10.1145/3613904.3642632}{doi:\nolinkurl{10.1145/3613904.3642632}}


\bibitem[Pavel et~al\mbox{.}(2017)]%
        {amy2017uistguidance}
\bibfield{author}{\bibinfo{person}{Amy Pavel}, \bibinfo{person}{Bj\"{o}rn Hartmann}, {and} \bibinfo{person}{Maneesh Agrawala}.} \bibinfo{year}{2017}\natexlab{}.
\newblock \showarticletitle{Shot Orientation Controls for Interactive Cinematography with 360 Video}. In \bibinfo{booktitle}{\emph{Proceedings of the 30th Annual ACM Symposium on User Interface Software and Technology}} (Qu\'{e}bec City, QC, Canada) \emph{(\bibinfo{series}{UIST '17})}. \bibinfo{publisher}{Association for Computing Machinery}, \bibinfo{address}{New York, NY, USA}, \bibinfo{pages}{289–297}.
\newblock
\showISBNx{9781450349819}
\href{https://doi.org/10.1145/3126594.3126636}{doi:\nolinkurl{10.1145/3126594.3126636}}


\bibitem[Pavel et~al\mbox{.}(2020)]%
        {amy2020rescribe}
\bibfield{author}{\bibinfo{person}{Amy Pavel}, \bibinfo{person}{Gabriel Reyes}, {and} \bibinfo{person}{Jeffrey~P. Bigham}.} \bibinfo{year}{2020}\natexlab{}.
\newblock \showarticletitle{Rescribe: Authoring and Automatically Editing Audio Descriptions}. In \bibinfo{booktitle}{\emph{Proceedings of the 33rd Annual ACM Symposium on User Interface Software and Technology}} (Virtual Event, USA) \emph{(\bibinfo{series}{UIST '20})}. \bibinfo{publisher}{Association for Computing Machinery}, \bibinfo{address}{New York, NY, USA}, \bibinfo{pages}{747–759}.
\newblock
\showISBNx{9781450375146}
\href{https://doi.org/10.1145/3379337.3415864}{doi:\nolinkurl{10.1145/3379337.3415864}}


\bibitem[Peng et~al\mbox{.}(2021)]%
        {peng2021slidecho}
\bibfield{author}{\bibinfo{person}{Yi-Hao Peng}, \bibinfo{person}{Jeffrey~P Bigham}, {and} \bibinfo{person}{Amy Pavel}.} \bibinfo{year}{2021}\natexlab{}.
\newblock \showarticletitle{Slidecho: Flexible non-visual exploration of presentation videos}. In \bibinfo{booktitle}{\emph{Proceedings of the 23rd International ACM SIGACCESS Conference on Computers and Accessibility}}. \bibinfo{pages}{1--12}.
\newblock


\bibitem[Reinders et~al\mbox{.}(2025)]%
        {reinders2025brought}
\bibfield{author}{\bibinfo{person}{Samuel Reinders}, \bibinfo{person}{Matthew Butler}, {and} \bibinfo{person}{Kim Marriott}.} \bibinfo{year}{2025}\natexlab{}.
\newblock \showarticletitle{" It Brought the Model to Life": Exploring the Embodiment of Multimodal I3Ms for People who are Blind or have Low Vision}. In \bibinfo{booktitle}{\emph{Proceedings of the 2025 CHI Conference on Human Factors in Computing Systems}}. \bibinfo{pages}{1--19}.
\newblock


\bibitem[Stangl et~al\mbox{.}(2023)]%
        {stangl2023potential}
\bibfield{author}{\bibinfo{person}{Abigale Stangl}, \bibinfo{person}{Shasta Ihorn}, \bibinfo{person}{Yue-Ting Siu}, \bibinfo{person}{Aditya Bodi}, \bibinfo{person}{Mar Castanon}, \bibinfo{person}{Lothar~D Narins}, {and} \bibinfo{person}{Ilmi Yoon}.} \bibinfo{year}{2023}\natexlab{}.
\newblock \showarticletitle{The potential of a visual dialogue agent in a tandem automated audio description system for videos}. In \bibinfo{booktitle}{\emph{Proceedings of the 25th International ACM SIGACCESS Conference on Computers and Accessibility}}. \bibinfo{pages}{1--17}.
\newblock


\bibitem[Stangl et~al\mbox{.}(2021)]%
        {stangl2021going}
\bibfield{author}{\bibinfo{person}{Abigale Stangl}, \bibinfo{person}{Nitin Verma}, \bibinfo{person}{Kenneth~R Fleischmann}, \bibinfo{person}{Meredith~Ringel Morris}, {and} \bibinfo{person}{Danna Gurari}.} \bibinfo{year}{2021}\natexlab{}.
\newblock \showarticletitle{Going beyond one-size-fits-all image descriptions to satisfy the information wants of people who are blind or have low vision}. In \bibinfo{booktitle}{\emph{Proceedings of the 23rd international ACM SIGACCESS conference on computers and accessibility}}. \bibinfo{pages}{1--15}.
\newblock


\bibitem[Van~Daele et~al\mbox{.}(2024)]%
        {van2024making}
\bibfield{author}{\bibinfo{person}{Tess Van~Daele}, \bibinfo{person}{Akhil Iyer}, \bibinfo{person}{Yuning Zhang}, \bibinfo{person}{Jalyn~C Derry}, \bibinfo{person}{Mina Huh}, {and} \bibinfo{person}{Amy Pavel}.} \bibinfo{year}{2024}\natexlab{}.
\newblock \showarticletitle{Making short-form videos accessible with hierarchical video summaries}. In \bibinfo{booktitle}{\emph{Proceedings of the 2024 CHI Conference on Human Factors in Computing Systems}}. \bibinfo{pages}{1--17}.
\newblock


\bibitem[Xu et~al\mbox{.}(2024)]%
        {xu2024memory}
\bibfield{author}{\bibinfo{person}{Shuchang Xu}, \bibinfo{person}{Chang Chen}, \bibinfo{person}{Zichen Liu}, \bibinfo{person}{Xiaofu Jin}, \bibinfo{person}{Lin-Ping Yuan}, \bibinfo{person}{Yukang Yan}, {and} \bibinfo{person}{Huamin Qu}.} \bibinfo{year}{2024}\natexlab{}.
\newblock \showarticletitle{Memory Reviver: Supporting Photo-Collection Reminiscence for People with Visual Impairment via a Proactive Chatbot}. In \bibinfo{booktitle}{\emph{Proceedings of the 37th Annual ACM Symposium on User Interface Software and Technology}} (Pittsburgh, PA, USA) \emph{(\bibinfo{series}{UIST '24})}. \bibinfo{publisher}{Association for Computing Machinery}, \bibinfo{address}{New York, NY, USA}, Article \bibinfo{articleno}{88}, \bibinfo{numpages}{17}~pages.
\newblock
\showISBNx{9798400706288}
\href{https://doi.org/10.1145/3654777.3676336}{doi:\nolinkurl{10.1145/3654777.3676336}}


\bibitem[Xu et~al\mbox{.}(2025a)]%
        {xu2025danmu}
\bibfield{author}{\bibinfo{person}{Shuchang Xu}, \bibinfo{person}{Xiaofu Jin}, \bibinfo{person}{Huamin Qu}, {and} \bibinfo{person}{Yukang Yan}.} \bibinfo{year}{2025}\natexlab{a}.
\newblock \showarticletitle{DanmuA11y: Making Time-Synced On-Screen Video Comments (Danmu) Accessible to Blind and Low Vision Users via Multi-Viewer Audio Discussions}. In \bibinfo{booktitle}{\emph{Proceedings of the 2025 CHI Conference on Human Factors in Computing Systems}} \emph{(\bibinfo{series}{CHI '25})}. \bibinfo{publisher}{Association for Computing Machinery}, \bibinfo{address}{New York, NY, USA}, Article \bibinfo{articleno}{293}, \bibinfo{numpages}{22}~pages.
\newblock
\showISBNx{9798400713941}
\href{https://doi.org/10.1145/3706598.3713496}{doi:\nolinkurl{10.1145/3706598.3713496}}


\bibitem[Xu et~al\mbox{.}(2025b)]%
        {xu2025branch}
\bibfield{author}{\bibinfo{person}{Shuchang Xu}, \bibinfo{person}{Xiaofu Jin}, \bibinfo{person}{Wenshuo Zhang}, \bibinfo{person}{Huamin Qu}, {and} \bibinfo{person}{Yukang Yan}.} \bibinfo{year}{2025}\natexlab{b}.
\newblock \showarticletitle{Branch Explorer: Leveraging Branching Narratives to Support Interactive 360° Video Viewing for Blind and Low Vision Users}. In \bibinfo{booktitle}{\emph{Proceedings of the 38th Annual ACM Symposium on User Interface Software and Technology}} \emph{(\bibinfo{series}{UIST '25})}. \bibinfo{pages}{1--18}.
\newblock


\end{thebibliography}

\appendix

\end{document}